# THE PROPERTIES OF THE FIRST EQUATION OF THE VLASOV CHAIN OF EQUATIONS

## E.E. Perepelkin[a], B.I. Sadovnikov[a], N.G. Inozemtseva[b]

[a] Faculty of Physics, Lomonosov Moscow State University, Moscow, 119991 Russia
e-mail: pevgeny@jinr.ru
[b] Dubna University, Moscow region, Moscow, 141980 Russia
e-mail: nginozv@mail.ru

**Annotation**

A mathematically rigorous derivation of the first Vlasov equation as a well-known Schrödinger equation for the probabilistic description of a system and families of the classic diffusion equations and heat conduction for the deterministic description of physical systems was inferred. A physical meaning of the phase of the wave function which is a scalar potential of the probabilistic flow velocity is demonstrated. Occurrence of the velocity potential vortex component leads to the Pauli equation for one of the spinar components. A scheme of the construction of the Schrödinger equation solving from the Vlasov equation solving and vice-versa is shown. A process of introduction of the potential to the Schrödinger equation and its interpretation are given. The analysis of the potential properties gives us the Maxwell equation, the equation of the kinematic point movement, and the movement of the medium within electromagnetic fields equation.

## Introduction

The main aim of statistical physics is study and explanation of the statistical mechanics laws proceeding from the first fundamental principles. Many of the equations and relationships describing various physical processes were often obtained and applied by the phenomenological method without a rigorous mathematical reasoning. A history of the developing of the quantum mechanics is an example of it.

The aim of physics is not only obtaining the equations that describe the processes but also the study of its properties, interrelation with other equations, and the analysis of the properties of these equations. We can see it by the example of the Painleve analysis, research of the differential equations group properties, and its properties with the help of Lax pair, methods of the inverse scattering theory and many others [1, 2].

This paper contains an analysis of the first equation of the Vlasov infinite chain of equations [3].The second equation of the Vlasov chain equations has been widely applied in accelerator physics, plasma physics, thermonuclear fusion problems, solid state physics, and crystals [4-6].

This paper is devoted to the first equation of the Vlasov chain of equations which is formally similar to the continuity equation, though it is written not for the classic subject density function but for the probability density function. Such "dualism" allows observing the equation both from the position of a classic – deterministic and probabilistic approach in the description of physical processes.

### 1. Phase as a scalar velocity potential

The first equation of the Vlasov chain of equations [3, 7] is written down for the probability density function $f_1(\vec{r},t)$ and has the following form:

$$\frac{\partial f_1(\vec{r},t)}{\partial t} + \text{div}_r \left[ f_1(\vec{r},t) \langle \vec{v} \rangle (\vec{r},t) \right] = 0 \tag{1}$$

where

$$\langle \vec{v} \rangle (\vec{r},t) = \frac{\int_{\Omega_v} f_2(\vec{r},\vec{v},t) \vec{v} d^3 v}{f_1(\vec{r},t)}. \tag{2}$$

Let us notice that Vlasov derived it (1) not for the particle system but for a particle the behaviour of which is characterized by the probability density function $f_1(\vec{r},t)$. On the other hand the equation (1) may be regarded as the conservation of mass or charge at the hydrodynamic approximation equation, which is derived for the medium and not for a particle.

In further computations the function $f_1(\vec{r},t)$ will be designated as $f(\vec{r},t)$, and the current density as $f_1(\vec{r},t) \langle \vec{v} \rangle (\vec{r},t) = \vec{J}(\vec{r},t)$. As the probability density function $f(\vec{r},t)$ is determined as a positive function, it can be presented as the unit square of a certain complex function $\Psi(\vec{r},t)$, i.e.:

$$f(\vec{r},t) = |\Psi(\vec{r},t)|^2 = |\overline{\Psi}(\vec{r},t)|^2 = \Psi(\vec{r},t) \overline{\Psi}(\vec{r},t) \geq 0. \tag{3}$$

Let us put expression (3) into equation (1) and we will derive the equation for the function $\Psi(\vec{r},t)$. It is important to note that $\Psi(\vec{r},t)$ is a complex-valued function of the real variables $\vec{r}$ and $t$. Substitution of (3) into (1) gives us differentiation of real variables $\vec{r}$ and $t$.

According to the Helmholtz theorem, let us consider the field of vectors $\vec{F}(\vec{r})$ as the superposition of fields $\vec{F}_1(\vec{r})$ and $\vec{F}_2(\vec{r})$ which satisfy the $\text{div}\,\vec{F}_1(\vec{r}) = 0$ and $\text{rot}\,\vec{F}_2(\vec{r}) = 0$ conditions. In that way, the vector field of average velocity (2) has the following form:

$$\langle \vec{v} \rangle (\vec{r},t) = -\alpha \nabla \Phi(\vec{r},t) + \gamma \vec{A}(\vec{r},t), \ \text{div}\,\vec{A} = 0, \tag{4}$$

where $\alpha, \gamma$ are certain real constants, $\Phi(\vec{r},t)$ – scalar potential of velocity (2), and $\vec{A}(\vec{r},t)$ – vortex component of velocity (2). Let us rewrite the form (4):

$$\begin{aligned}\langle \vec{v} \rangle (\vec{r},t) &= -\alpha \nabla \Phi(\vec{r},t) + \gamma \vec{A}(\vec{r},t) = i^2 \alpha \nabla \Phi + \gamma \vec{A} = \\ &= i\alpha \nabla (0 + i\Phi) + \gamma \vec{A} = i\alpha \nabla \left( \ln \left| \frac{\Psi}{\overline{\Psi}} \right| + i\Phi \right) + \gamma \vec{A}.\end{aligned} \tag{5}$$

As the function $\Psi(\vec{r},t)$ is complex, it satisfies the exponential form of expression with the form of:

$$\Psi(\vec{r},t) = |\Psi(\vec{r},t)| e^{i\varphi(\vec{r},t)}, \tag{6}$$

where $\varphi(\vec{r},t)$ is the phase. Taking into consideration expression (6) the $\dfrac{\Psi}{\overline{\Psi}}$ function will have the form:

$$\frac{\Psi(\vec{r},t)}{\overline{\Psi}(\vec{r},t)} = e^{i2\varphi(\vec{r},t)} \Rightarrow \mathrm{Arg}\left[\frac{\Psi(\vec{r},t)}{\overline{\Psi}(\vec{r},t)}\right] = 2\varphi(\vec{r},t) + 2\pi k \stackrel{\mathrm{det}}{=} \Phi(\vec{r},t). \qquad (7)$$

So the velocity potential (2) in (7) was determined as a complex function argument $\dfrac{\Psi}{\overline{\Psi}}$. Considering the notation (7), we will rewrite the form (5):

$$\langle \vec{v} \rangle(\vec{r},t) = i\alpha \nabla \left( \ln\left|\frac{\Psi}{\overline{\Psi}}\right| + i\,\mathrm{Arg}\left[\frac{\Psi}{\overline{\Psi}}\right] \right) + \gamma \vec{A} = i\alpha \nabla \mathrm{Ln}\left[\frac{\Psi}{\overline{\Psi}}\right] + \gamma \vec{A} =$$
$$= i\alpha \nabla \left[ \mathrm{Ln}(\Psi) - \mathrm{Ln}(\overline{\Psi}) \right] + \gamma \vec{A} = i\alpha \left[ \frac{\nabla \Psi}{\Psi} - \frac{\nabla \overline{\Psi}}{\overline{\Psi}} \right] + \gamma \vec{A}. \qquad (8)$$

## 2. Parabolic equation

Considering (3) and (8), let us rewrite equation (1)

$$\overline{\Psi}\frac{\partial \Psi}{\partial t} + \Psi \frac{\partial \overline{\Psi}}{\partial t} + i\alpha\,\mathrm{div}_r\left[ \Psi\overline{\Psi}\left( \frac{\nabla \Psi}{\Psi} - \frac{\nabla \overline{\Psi}}{\overline{\Psi}} - i\frac{\gamma}{\alpha}\vec{A} \right) \right] = 0,$$

or

$$\overline{\Psi}\frac{\partial \Psi}{\partial t} + \Psi \frac{\partial \overline{\Psi}}{\partial t} + i\alpha\,\mathrm{div}_r\left[ \overline{\Psi}\nabla \Psi - \Psi \nabla \overline{\Psi} - i\frac{\gamma}{\alpha} \Psi \overline{\Psi} \vec{A} \right] = 0. \qquad (9)$$

Now we intend to write out the expressions $\mathrm{div}_r\left[\overline{\Psi}\nabla\Psi - \Psi\nabla\overline{\Psi}\right]$ and $\mathrm{div}_r\left[\Psi\overline{\Psi}\vec{A}\right]$:

$$\mathrm{div}_r\left[\overline{\Psi}\nabla\Psi - \Psi\nabla\overline{\Psi}\right] = \nabla\overline{\Psi}\nabla\Psi + \overline{\Psi}\Delta\Psi - \nabla\Psi\nabla\overline{\Psi} - \Psi\Delta\overline{\Psi} = \overline{\Psi}\Delta\Psi - \Psi\Delta\overline{\Psi}, \qquad (10)$$
$$\mathrm{div}_r\left[\Psi\overline{\Psi}\vec{A}\right] = (\nabla\Psi\overline{\Psi}, \vec{A}) + \Psi\overline{\Psi}(\nabla, \vec{A}) = \overline{\Psi}(\vec{A}, \nabla\Psi) + \Psi(\vec{A}, \nabla\overline{\Psi}).$$

Expression (4) is considered in (10), $\mathrm{div}\,\vec{A} = 0$. Let us substitute (10) into (9):

$$\overline{\Psi}\frac{\partial \Psi}{\partial t} + \Psi \frac{\partial \overline{\Psi}}{\partial t} + i\alpha\left(\overline{\Psi}\Delta\Psi - \Psi\Delta\overline{\Psi}\right) + \overline{\Psi}\left(\gamma\vec{A}, \nabla\Psi\right) + \Psi\left(\gamma\vec{A}, \nabla\overline{\Psi}\right) = 0. \qquad (11)$$

Now let us group the terms of expression (11):

$$\overline{\Psi}\left[\frac{\partial \Psi}{\partial t} + i\alpha\Delta\Psi + \gamma(\vec{A}, \nabla\Psi)\right] + \Psi\left[\frac{\partial \overline{\Psi}}{\partial t} - i\alpha\Delta\overline{\Psi} + \gamma(\vec{A}, \nabla\overline{\Psi})\right] = 0. \qquad (12)$$

Here we introduce the notation of the differential operator:

$$\hat{p} \stackrel{\text{det}}{=} -\frac{i}{\beta}\nabla, \ \bar{\hat{p}} = \frac{i}{\beta}\nabla, \ \hat{p}^2 = \bar{\hat{p}}^2 = -\frac{1}{\beta^2}\Delta,$$
$$\nabla = i\beta\hat{p}, \ \nabla = -i\beta\bar{\hat{p}}, \ \Delta = -\beta^2\hat{p}^2 = -\beta^2\bar{\hat{p}}^2,$$
(13)

where $\beta \in \mathbb{R}$, $\beta \neq 0$. Then the expression (12) will have the form:

$$\bar{\Psi}\left[\frac{1}{\beta}\frac{\partial}{\partial t} - i\alpha\beta\left(\hat{p}^2 - \frac{\gamma}{\alpha\beta}(\vec{A},\hat{p})\right)\right]\Psi + \Psi\left[\frac{1}{\beta}\frac{\partial}{\partial t} + i\alpha\beta\left(\bar{\hat{p}}^2 - \frac{\gamma}{\alpha\beta}(\vec{A},\bar{\hat{p}})\right)\right]\bar{\Psi} = 0. \quad (14)$$

Expression (14) may be written in another way, provided :

$$\hat{p}^2 - \frac{\gamma}{\alpha\beta}(\vec{A},\hat{p}) = \left(\hat{p} - \frac{\gamma}{2\alpha\beta}\vec{A}\right)^2 - \frac{\gamma^2}{4\alpha^2\beta^2}|\vec{A}|^2,$$
$$\bar{\hat{p}}^2 - \frac{\gamma}{\alpha\beta}(\vec{A},\bar{\hat{p}}) = \left(\bar{\hat{p}} - \frac{\gamma}{2\alpha\beta}\vec{A}\right)^2 - \frac{\gamma^2}{4\alpha^2\beta^2}|\vec{A}|^2,$$
(15)

as

$$\left(\hat{p} - \frac{\gamma}{2\alpha\beta}\vec{A}\right)^2\Psi = \left(\hat{p} - \frac{\gamma}{2\alpha\beta}\vec{A}\right)\left(\hat{p} - \frac{\gamma}{2\alpha\beta}\vec{A}\right)\Psi = \left(\hat{p} - \frac{\gamma}{2\alpha\beta}\vec{A}\right)\left(\hat{p}\Psi - \frac{\gamma}{2\alpha\beta}\vec{A}\Psi\right) =$$
$$= \hat{p}^2\Psi - \frac{\gamma}{2\alpha\beta}(\hat{p},\vec{A}\Psi) - \frac{\gamma}{2\alpha\beta}(\vec{A},\hat{p}\Psi) + \frac{\gamma^2}{4\alpha^2\beta^2}|\vec{A}|^2\Psi =$$
$$= \hat{p}^2\Psi - \frac{\gamma}{2\alpha\beta}\Psi(\hat{p},\vec{A}) - \frac{\gamma}{2\alpha\beta}(\vec{A},\hat{p}\Psi) - \frac{\gamma}{2\alpha\beta}(\vec{A},\hat{p}\Psi) + \frac{\gamma^2}{4\alpha^2\beta^2}|\vec{A}|^2 =$$
$$= \hat{p}^2\Psi - \frac{\gamma}{\alpha\beta}(\vec{A},\hat{p})\Psi + \frac{\gamma^2}{4\alpha^2\beta^2}|\vec{A}|^2\Psi.$$

Expression (14) with the consideration of the (15) will have the form:

$$\bar{\Psi}\left[\frac{1}{\beta}\frac{\partial}{\partial t} - i\alpha\beta\left(\hat{p} - \frac{\gamma}{2\alpha\beta}\vec{A}\right)^2 + i\frac{\gamma^2}{4\alpha\beta}|\vec{A}|^2\right]\Psi + \Psi\left[\frac{1}{\beta}\frac{\partial}{\partial t} + i\alpha\beta\left(\bar{\hat{p}} - \frac{\gamma}{2\alpha\beta}\vec{A}\right)^2 - i\frac{\gamma^2}{4\alpha\beta}|\vec{A}|^2\right]\bar{\Psi} = 0,$$

$$\bar{\Psi}\left[\frac{1}{\beta}\frac{\partial}{\partial t} - i\alpha\beta\left(\hat{p} - \frac{\gamma}{2\alpha\beta}\vec{A}\right)^2\right]\Psi + \Psi\left[\frac{1}{\beta}\frac{\partial}{\partial t} + i\alpha\beta\left(\bar{\hat{p}} - \frac{\gamma}{2\alpha\beta}\vec{A}\right)^2\right]\bar{\Psi} = 0. \quad (16)$$

According to that, we can work both at expression (14) and (16). Now we intend to consider the expression (14). There are two possible variants of converting it. Let us deal with each of them separately. Here is notation of the linear operator $L$:

$$L = \frac{1}{\beta}\frac{\partial}{\partial t} - i\alpha\beta\left(\hat{p}^2 - \frac{\gamma}{\alpha\beta}(\vec{A},\hat{p})\right), \ \bar{L} = \frac{1}{\beta}\frac{\partial}{\partial t} + i\alpha\beta\left(\bar{\hat{p}}^2 - \frac{\gamma}{\alpha\beta}(\vec{A},\bar{\hat{p}})\right), \quad (17)$$

Then expression (14) with the consideration of the introduced notation (17) will be:

$$\bar{\Psi}L\Psi + \Psi\bar{L}\bar{\Psi} = 0, \qquad (18)$$

or

$$\Lambda + \bar{\Lambda} = 0,$$

where $\Lambda = \bar{\Psi}L\Psi$. Expression (18) means that $\operatorname{Re}\Lambda = 0$. If the real part is equal to zero then $\Lambda$ is a virtual value, i.e.:

$$\Lambda = iv, \; v \in \mathbb{R},$$

or

$$\bar{\Psi}L\Psi = iv, \; L\Psi = i\frac{v}{\bar{\Psi}} = i\frac{v}{\bar{\Psi}}\frac{\Psi}{\Psi} = i\frac{v}{|\Psi|^2}\Psi = i\mu\Psi, \mu \in \mathbb{R},$$

$$L\Psi = i\mu\Psi. \qquad (19)$$

It is important to note that the $L$ operator is anti-Hermitian. We consider the dot product at the Hilbert space above the field of complex numbers, and use (18):

$$\langle \Psi, L\Psi \rangle = \int \Psi \bar{L}\bar{\Psi} d\omega = -\int \bar{\Psi}L\Psi d\omega = -\int (L\Psi)\bar{\Psi} d\omega = -\langle L\Psi, \Psi \rangle,$$

or

$$\langle \Psi, L\Psi \rangle = \langle \bar{L}\Psi, \Psi \rangle \Rightarrow \langle \bar{L}\Psi, \Psi \rangle = -\langle L\Psi, \Psi \rangle. \qquad (20)$$

It is well-known that an anti-Hermitian operator has pure imaginary eigenvalue that matches expression (19).

Let us return to expression (14) and divide it by $i$ value:

$$\bar{\Psi}\left[-\frac{i}{\beta}\frac{\partial}{\partial t} - \alpha\beta\left(\hat{p}^2 - \frac{\gamma}{\alpha\beta}(\vec{A},\hat{p})\right)\right]\Psi + \Psi\left[-\frac{i}{\beta}\frac{\partial}{\partial t} + \alpha\beta\left(\bar{\hat{p}}^2 - \frac{\gamma}{\alpha\beta}(\vec{A},\bar{\hat{p}})\right)\right]\bar{\Psi} = 0.$$

or

$$\bar{\Psi}\left[-\frac{i}{\beta}\frac{\partial}{\partial t} - \alpha\beta\left(\hat{p}^2 - \frac{\gamma}{\alpha\beta}(\vec{A},\hat{p})\right)\right]\Psi - \Psi\left[\frac{i}{\beta}\frac{\partial}{\partial t} - \alpha\beta\left(\bar{\hat{p}}^2 - \frac{\gamma}{\alpha\beta}(\vec{A},\bar{\hat{p}})\right)\right]\bar{\Psi} = 0. \qquad (21)$$

Here is the notation of the linear operator $\mathcal{L}$ as:

$$\mathcal{L} = -\frac{i}{\beta}\frac{\partial}{\partial t} - \alpha\beta\left(\hat{p}^2 - \frac{\gamma}{\alpha\beta}(\vec{A},\hat{p})\right), \; \bar{\mathcal{L}} = \frac{i}{\beta}\frac{\partial}{\partial t} - \alpha\beta\left(\bar{\hat{p}}^2 - \frac{\gamma}{\alpha\beta}(\vec{A},\bar{\hat{p}})\right). \qquad (22)$$

Using the notation (22), equation (21) will be the following:

$$\bar{\Psi}\mathcal{L}\Psi - \Psi\bar{\mathcal{L}}\bar{\Psi} = 0,$$

or

$$M - \bar{M} = 0, \qquad (23)$$

where $M = \bar{\Psi}\mathcal{L}\Psi$. Expression (23) means that $\operatorname{Im} M = 0$. If the imaginary part is equal to zero, then $M$ is the real value, i.e.:

$$M = u, \ u \in \mathbb{R},$$

or

$$\bar{\Psi}\mathcal{L}\Psi = u, \ \mathcal{L}\Psi = \frac{u}{\bar{\Psi}} = \frac{u}{\bar{\Psi}}\frac{\Psi}{\Psi} = \frac{u}{|\Psi|^2}\Psi = -U\Psi, \ U(\vec{r},t) \in \mathbb{R},$$

$$\mathcal{L}\Psi = -U\Psi. \tag{24}$$

It is significant to note that the $\mathcal{L}$ operator is Hermitian. We consider the dot product at the Hilbert space above the field of complex numbers, and use (23):

$$\langle \Psi, \mathcal{L}\Psi \rangle = \int \Psi \overline{\mathcal{L}\Psi} d\omega = \int \bar{\Psi}\mathcal{L}\Psi d\omega = \int (\mathcal{L}\Psi)\bar{\Psi} d\omega = \langle \mathcal{L}\Psi, \Psi \rangle,$$

i.e.

$$\langle \Psi, \mathcal{L}\Psi \rangle = \langle \bar{\mathcal{L}}\Psi, \Psi \rangle \Rightarrow \langle \bar{\mathcal{L}}\Psi, \Psi \rangle = \langle \mathcal{L}\Psi, \Psi \rangle. \tag{25}$$

From the comparison of the operators $L$ and $\mathcal{L}$ (17) and (23) determinations we can see that:

$$L\Psi = i\mathcal{L}\Psi. \tag{26}$$

Taking expressions (19) and (24) into account, equation (22) will have the form:

$$i\mu\Psi = -iU\Psi,$$

from this

$$\mu = -U. \tag{27}$$

As a result, equation (19) and (24) considering (27) will be the following:

$$L\Psi + iU\Psi = \frac{1}{\beta}\frac{\partial \Psi}{\partial t} - i\alpha\beta\left(\hat{p}^2 - \frac{\gamma}{\alpha\beta}(\vec{A},\hat{p})\right)\Psi + iU\Psi = 0,$$

$$\mathcal{L}\Psi + U\Psi = -\frac{i}{\beta}\frac{\partial}{\partial t} - \alpha\beta\left(\hat{p}^2 - \frac{\gamma}{\alpha\beta}(\vec{A},\hat{p})\right) + U\Psi = 0.$$

or

$$\frac{i}{\beta}\frac{\partial \Psi}{\partial t} = -\alpha\beta\left(\hat{p}^2 - \frac{\gamma}{\alpha\beta}(\vec{A},\hat{p})\right)\Psi + U\Psi. \tag{28}$$

Considering (15), equation (28) is:

$$\frac{i}{\beta}\frac{\partial \Psi}{\partial t} = -\alpha\beta\left(\hat{p} - \frac{\gamma}{2\alpha\beta}\vec{A}\right)^2 + \frac{1}{2\alpha\beta}\frac{|\gamma\vec{A}|^2}{2}\Psi + U\Psi \tag{29}$$

or

$$\frac{i}{\beta}\frac{\partial \Psi}{\partial t} = -\alpha\beta\left(\hat{p} - \frac{\gamma}{2\alpha\beta}\vec{A}\right)^2 - L\Psi,$$

where

$$L \stackrel{det}{=} -\frac{1}{2\alpha\beta}\frac{\left|\gamma\vec{A}\right|^2}{2} - U = T_{rot} - U.$$

So, expression (14) resulted the deduction of equation (29). Let us derive an equation for expression (16). Again, introduce the operators:

$$L_2 = \frac{1}{\beta}\frac{\partial}{\partial t} - i\alpha\beta\left(\hat{p} - \frac{\gamma}{2\alpha\beta}\vec{A}\right)^2, \quad \bar{L}_2 = \frac{1}{\beta}\frac{\partial}{\partial t} + i\alpha\beta\left(\hat{\bar{p}} - \frac{\gamma}{2\alpha\beta}\vec{A}\right)^2, \quad (30)$$

so that expression (16) will take a form with the consider of the introduced notation of (30):

$$\bar{\Psi}L_2\Psi + \Psi\bar{L}_2\bar{\Psi} = 0, \quad (31)$$

or

$$\Lambda_2 + \bar{\Lambda}_2 = 0,$$

where $\Lambda_2 = \bar{\Psi}L_2\Psi$. Expression (31) means that $\text{Re}\,\Lambda_2 = 0$. If the real part is equal to zero then $\Lambda_2$ is a virtual value, i.e.:

$$\Lambda_2 = iv_2, \ v_2 \in \mathbb{R},$$

or

$$\bar{\Psi}L_2\Psi = iv_2, \ L_2\Psi = i\frac{v_2}{\bar{\Psi}} = i\frac{v_2}{\bar{\Psi}}\frac{\Psi}{\Psi} = i\frac{v_2}{|\Psi|^2}\Psi = i\mu_2\Psi, \mu_2 \in \mathbb{R},$$

$$L\Psi = i\mu\Psi. \quad (32)$$

It is important to note that the $L_2$ operator is anti-Hermitian. We consider the dot product at the Hilbert space above the field of complex numbers, and use (31):

$$\langle \Psi, L_2\Psi \rangle = \int \Psi \bar{L}_2 \bar{\Psi} d\omega = -\int \bar{\Psi} L_2 \Psi d\omega = -\int (L_2\Psi)\bar{\Psi} d\omega = -\langle L_2\Psi, \Psi \rangle,$$

i.e.

$$\langle \Psi, L_2\Psi \rangle = \langle \bar{L}_2\Psi, \Psi \rangle \Rightarrow \langle \bar{L}_2\Psi, \Psi \rangle = -\langle L_2\Psi, \Psi \rangle. \quad (33)$$

It is well-known that an anti-Hermitian operator has pure imaginary eigenvalue that matches expression (32).

Let us return to expression (16) and divide it by $i$ value:

$$\bar{\Psi}\left[-\frac{i}{\beta}\frac{\partial}{\partial t} - \alpha\beta\left(\hat{p} - \frac{\gamma}{2\alpha\beta}\vec{A}\right)^2\right]\Psi + \Psi\left[-\frac{i}{\beta}\frac{\partial}{\partial t} + \alpha\beta\left(\hat{\bar{p}} - \frac{\gamma}{2\alpha\beta}\vec{A}\right)^2\right]\bar{\Psi} = 0$$

or

$$\bar{\Psi}\left[-\frac{i}{\beta}\frac{\partial}{\partial t}-\alpha\beta\left(\hat{p}-\frac{\gamma}{2\alpha\beta}\vec{A}\right)^2\right]\Psi-\Psi\left[\frac{i}{\beta}\frac{\partial}{\partial t}-\alpha\beta\left(\bar{\hat{p}}-\frac{\gamma}{2\alpha\beta}\vec{A}\right)^2\right]\bar{\Psi}=0. \quad (34)$$

The linear operator $\mathcal{L}_2$ will be marked as:

$$\mathcal{L}_2=-\frac{i}{\beta}\frac{\partial}{\partial t}-\alpha\beta\left(\hat{p}-\frac{\gamma}{2\alpha\beta}\vec{A}\right)^2, \quad \bar{\mathcal{L}}_2=\frac{i}{\beta}\frac{\partial}{\partial t}-\alpha\beta\left(\bar{\hat{p}}-\frac{\gamma}{2\alpha\beta}\vec{A}\right)^2. \quad (35)$$

Using the notation (35), equation (34) will be the following:

$$\bar{\Psi}\mathcal{L}_2\Psi-\Psi\bar{\mathcal{L}}_2\bar{\Psi}=0,$$

or

$$M_2-\bar{M}_2=0, \quad (36)$$

where $M_2=\bar{\Psi}\mathcal{L}_2\Psi$. Expression (36) means that $\mathrm{Im}\,M_2=0$. If the real part is equal to zero then $M_2$ is a virtual value, i.e.:

$$M_2=u_2,\ u_2\in\mathbb{R},$$

or

$$\bar{\Psi}\mathcal{L}_2\Psi=u_2,\ \mathcal{L}_2\Psi=\frac{u_2}{\bar{\Psi}}=\frac{u_2}{\bar{\Psi}}\frac{\Psi}{\Psi}=\frac{u_2}{|\Psi|^2}\Psi=-U_2\Psi,\ U_2(\vec{r},t)\in\mathbb{R},$$

$$\mathcal{L}_2\Psi=-U_2\Psi. \quad (37)$$

It is significant to note that the $\mathcal{L}_2$ operator is Hermitian. We consider the dot product at the Hilbert space above the field of complex numbers, and use (36):

$$\langle\Psi,\mathcal{L}_2\Psi\rangle=\int\Psi\bar{\mathcal{L}}_2\bar{\Psi}d\omega=\int\bar{\Psi}\mathcal{L}_2\Psi d\omega=\int(\mathcal{L}_2\Psi)\bar{\Psi}d\omega=\langle\mathcal{L}_2\Psi,\Psi\rangle,$$

i.e.

$$\langle\Psi,\mathcal{L}_2\Psi\rangle=\langle\bar{\mathcal{L}}_2\Psi,\Psi\rangle\Rightarrow\langle\bar{\mathcal{L}}_2\Psi,\Psi\rangle=\langle\mathcal{L}_2\Psi,\Psi\rangle. \quad (38)$$

From the comparison of the operators $L_2$ and $\mathcal{L}_2$ (30) and (35) determinations we can see that:

$$L_2\Psi=i\mathcal{L}_2\Psi. \quad (39)$$

Taking expressions (32) and (37) into account, equation (39) will have the form:

$$i\mu_2\Psi=-iU_2\Psi,$$

from this

$$\mu_2 = -U_2. \tag{40}$$

As a result, equation (32) and (37) considering (40) will be the following:

$$L_2\Psi + iU_2\Psi = \frac{1}{\beta}\frac{\partial \Psi}{\partial t} - i\alpha\beta\left(\hat{p} - \frac{\gamma}{2\alpha\beta}\vec{A}\right)^2\Psi + iU_2\Psi = 0,$$

$$\mathcal{L}_2\Psi + U_2\Psi = -\frac{i}{\beta}\frac{\partial}{\partial t} - \alpha\beta\left(\hat{p} - \frac{\gamma}{2\alpha\beta}\vec{A}\right)^2 + U_2\Psi = 0.$$

or

$$\frac{i}{\beta}\frac{\partial \Psi}{\partial t} = -\alpha\beta\left(\hat{p} - \frac{\gamma}{2\alpha\beta}\vec{A}\right)^2\Psi + U_2\Psi. \tag{41}$$

It is seen that the derived equation (41) differs from equation (29) in absence of the term $\frac{1}{2\alpha\beta}\frac{|\gamma\vec{A}|^2}{2}\Psi$. Not limiting the generalities one can consider that the function $U_2$ in equation (41) may keep the given addend, i.e.:

$$U_2 = U + \frac{1}{2\alpha\beta}\frac{|\gamma\vec{A}|^2}{2} = -\mathrm{L} \tag{42}$$

In this case, equation (41) grades into (29). If $\langle\vec{v}\rangle(\vec{r},t)$ is known, then we can find $f(\vec{r},t)$ from equation (1), and also the velocity potential $\Phi(\vec{r},t)$ and $\vec{A}(\vec{r},t)$, solving the equations:

$$\Delta\Phi(\vec{r},t) = -\frac{1}{\alpha}\operatorname{div}\langle\vec{v}\rangle(\vec{r},t) \stackrel{\det}{=} -g(\vec{r},t),$$
$$\operatorname{rot}\vec{A}(\vec{r},t) = \frac{1}{\gamma}\operatorname{rot}\langle\vec{v}\rangle(\vec{r},t) \stackrel{\det}{=} \vec{B}(\vec{r},t), \tag{43}$$

which are derived from the (4). The velocity potential $\Phi(\vec{r},t)$ within a constant value is equal to the phase $\varphi(\vec{r},t)$, according to the formula (7). As a result, we will obtain the function $\Psi$ in the form of

$$\Psi(\vec{r},t) = \sqrt{f(\vec{r},t)}e^{i\varphi(\vec{r},t)} = (-1)^k\sqrt{f(\vec{r},t)}e^{i\frac{\Phi(\vec{r},t)}{2}}. \tag{44}$$

From knowledge of $\Psi$, it is possible to determine the corresponding function $U$ by formula (24):

$$U(\vec{r},t) = -\frac{\overline{\Psi}\mathcal{L}\Psi}{f(\vec{r},t)}. \tag{45}$$

Let us transform the expression (45).

$$\Delta\Psi = \left(\nabla, e^{i\varphi}\nabla|\Psi|\right) + i\left(\nabla, e^{i\varphi}|\Psi|\nabla\varphi\right) = ie^{i\varphi}\left(\nabla\varphi, \nabla|\Psi|\right) + e^{i\varphi}\Delta|\Psi| - e^{i\varphi}|\Psi|(\nabla\varphi, \nabla\varphi) +$$

$$+ ie^{i\varphi}\left(\nabla|\Psi|, \nabla\varphi\right) + ie^{i\varphi}|\Psi|\Delta\varphi = e^{i\varphi}\left\{\Delta|\Psi| - |\Psi||\nabla\varphi|^2 + i\left[2(\nabla\varphi, \nabla|\Psi|) + |\Psi|\Delta\varphi\right]\right\},$$

$$\bar{\Psi}\mathcal{L}\Psi = -\frac{i}{\beta}\bar{\Psi}\frac{\partial\Psi}{\partial t} + \frac{\alpha}{\beta}\bar{\Psi}\Delta\Psi - i\frac{\gamma}{\beta}\bar{\Psi}\left(\vec{A}, \nabla\Psi\right) =$$

$$= -\frac{i}{\beta}|\Psi|\frac{\partial|\Psi|}{\partial t} + \frac{1}{\beta}|\Psi|^2\frac{\partial\varphi}{\partial t} + \frac{\alpha}{\beta}|\Psi|\left\{\Delta|\Psi| - |\Psi||\nabla\varphi|^2 + i\left[2(\nabla\varphi, \nabla|\Psi|) + |\Psi|\Delta\varphi\right]\right\} -$$

$$- i\frac{\gamma}{\beta}|\Psi|\left(\vec{A}, \nabla|\Psi|\right) + \frac{\gamma}{\beta}|\Psi|^2\left(\vec{A}, \nabla\varphi\right),$$

$$U = -\frac{\bar{\Psi}\mathcal{L}\Psi}{|\Psi|^2} =$$

$$= -\frac{1}{\beta}\frac{\partial\varphi}{\partial t} - \frac{\alpha}{\beta}\frac{\Delta|\Psi|}{|\Psi|} + \frac{\alpha}{\beta}|\nabla\varphi|^2 + \qquad\qquad (46)$$

$$+ i\frac{1}{|\Psi|}\left[\frac{1}{\beta}\frac{\partial|\Psi|}{\partial t} - \frac{\alpha}{\beta}2(\nabla\varphi, \nabla|\Psi|) - \frac{\alpha}{\beta}|\Psi|\Delta\varphi\right] + i\frac{\gamma}{\beta}\left(\vec{A}, \frac{\nabla|\Psi|}{|\Psi|}\right) - \frac{\gamma}{\beta}(\vec{A}, \nabla\varphi).$$

According to the determination (24) function $U$ is substantial, therefore the imaginary part of expression (46) is to become zero. Let us verify that. Indeed, considering expressions (3), (4), (7) we obtain the following:

$$\mathrm{Im}\,U = \frac{1}{|\Psi|^2}\left[\frac{1}{\beta}|\Psi|\frac{\partial|\Psi|}{\partial t} - \frac{\alpha}{\beta}2(\nabla\varphi, |\Psi|\nabla|\Psi|) - \frac{\alpha}{\beta}|\Psi|^2\Delta\varphi + \frac{\gamma}{\beta}|\Psi|(\vec{A}, \nabla|\Psi|)\right] =$$

$$= \frac{1}{|\Psi|^2}\left[\frac{1}{2\beta}\frac{\partial|\Psi|^2}{\partial t} - \frac{\alpha}{\beta}(\nabla\varphi, \nabla|\Psi|^2) - \frac{\alpha}{\beta}|\Psi|^2\Delta\varphi + \frac{\gamma}{2\beta}(\vec{A}, \nabla|\Psi|^2)\right] =$$

$$= \frac{1}{f}\left[\frac{1}{2\beta}\frac{\partial f}{\partial t} - \frac{\alpha}{\beta}(\nabla\varphi, \nabla f) - \frac{\alpha}{\beta}f\Delta\varphi + \frac{\gamma}{2\beta}(\vec{A}, \nabla f)\right] =$$

$$= \frac{1}{2\beta f}\left[\frac{\partial f}{\partial t} + \left(-2\alpha\nabla\varphi + \gamma\vec{A}, \nabla f\right) + f\left(\nabla, -2\alpha\nabla\varphi + \gamma\vec{A}\right)\right] =$$

$$= \frac{1}{2\beta f}\left[\frac{\partial f}{\partial t} + \left(\langle\vec{v}\rangle, \nabla f\right) + f\,\mathrm{div}\,\langle\vec{v}\rangle\right] = \frac{1}{2\beta f}\left\{\frac{\partial f}{\partial t} + \mathrm{div}\left[\langle\vec{v}\rangle f\right]\right\} = 0.$$

As a result, we obtain the notion for the $U$ function:

$$U(\vec{r},t) = -\frac{1}{\beta}\left\{\frac{\partial\varphi(\vec{r},t)}{\partial t} + \alpha\left[\frac{\Delta\sqrt{f(\vec{r},t)}}{\sqrt{f(\vec{r},t)}} - |\nabla\varphi(\vec{r},t)|^2\right] + \gamma(\vec{A}, \nabla\varphi)\right\}. \qquad (47)$$

According to that, expressions (43), (44) and (47) determine the correspondence between the original equation (1) and the new one (29). Knowing the solutions of equation (1), one can obtain the solutions of equation (29) and vice versa. Knowing the solutions of equation (29), one can obtain the solutions of equation (1) and vice versa.

### 3. The equation of the kinematic point motion

On the basis of the equation (1) one can derive the equation of motion both for kinematic points and for center of mass of the medium determined by the cumulative distribution function $f(\vec{r},t)$.

Let us derive the expression for the gradient of function $U(\vec{r},t)$. Considering expressions (4) and (7), potential gradient $\nabla \varphi(\vec{r},t)$ will be:

$$\nabla \varphi(\vec{r},t) = \frac{1}{2}\nabla \Phi(\vec{r},t) = -\frac{1}{2\alpha}\langle \vec{v}\rangle(\vec{r},t) + \frac{\gamma}{2\alpha}\vec{A}(\vec{r},t),$$

$$\nabla \frac{\partial}{\partial t}\varphi(\vec{r},t) = -\frac{1}{2\alpha}\frac{\partial}{\partial t}\left[\langle \vec{v}\rangle(\vec{r},t) - \gamma \vec{A}(\vec{r},t)\right], \qquad (48)$$

$$|\nabla \varphi|^2 = \frac{1}{4\alpha^2}\left(\langle \vec{v}\rangle - \gamma \vec{A},\langle \vec{v}\rangle - \gamma \vec{A}\right) = \frac{1}{4\alpha^2}\left[(\langle \vec{v}\rangle,\langle \vec{v}\rangle) - 2\gamma(\langle \vec{v}\rangle,\vec{A}) + \gamma^2(\vec{A},\vec{A})\right] =$$

$$= \frac{1}{4\alpha^2}\left[|\langle \vec{v}\rangle|^2 - 2\gamma(\langle \vec{v}\rangle,\vec{A}) + \gamma^2|\vec{A}|^2\right].$$

Using (48), the gradient of $U$ function (47) will be:

$$\nabla U = -\frac{1}{\beta}\left\{-\frac{1}{2\alpha}\frac{\partial}{\partial t}\left[\langle \vec{v}\rangle - \gamma \vec{A}\right] + \alpha \nabla \left[\frac{\Delta \sqrt{f}}{\sqrt{f}} - \frac{1}{4\alpha^2}|\langle \vec{v}\rangle|^2\right] + \frac{\gamma}{2\alpha}\nabla(\vec{A},\langle \vec{v}\rangle) - \right.$$

$$\left. -\frac{\gamma^2}{4\alpha}\nabla|\vec{A}|^2 - \frac{\gamma}{2\alpha}\nabla(\vec{A},\langle \vec{v}\rangle - \gamma \vec{A})\right\} = \qquad (49)$$

$$= -\frac{1}{\beta}\left\{-\frac{1}{2\alpha}\frac{\partial}{\partial t}\left[\langle \vec{v}\rangle - \gamma \vec{A}\right] + \alpha \nabla \left[\frac{\Delta \sqrt{f}}{\sqrt{f}} - \frac{1}{4\alpha^2}|\langle \vec{v}\rangle|^2\right] + \frac{\gamma}{2\alpha}\nabla(\vec{A},\langle \vec{v}\rangle) - \right.$$

$$\left. -\frac{\gamma^2}{4\alpha}\nabla|\vec{A}|^2 - \frac{\gamma}{2\alpha}\nabla(\vec{A},\langle \vec{v}\rangle) + \frac{\gamma^2}{2\alpha}\nabla|\vec{A}|^2\right\} =$$

$$= -\frac{1}{\beta}\left\{-\frac{1}{2\alpha}\frac{\partial}{\partial t}\left[\langle \vec{v}\rangle - \gamma \vec{A}\right] + \alpha \nabla \left[\frac{\Delta \sqrt{f}}{\sqrt{f}} - \frac{1}{4\alpha^2}|\langle \vec{v}\rangle|^2\right] + \frac{\gamma^2}{4\alpha}\nabla|\vec{A}|^2\right\}$$

Now take the property (4) of the vector field $\langle \vec{v}\rangle$ into account, i.e.

$$\text{rot}\langle \vec{v}\rangle = \gamma\, \text{rot}\,\vec{A}. \qquad (50)$$

(50) implies that

$$\left[\langle \vec{v}\rangle, \gamma\,\text{rot}\,\vec{A}\right] = \left[\langle \vec{v}\rangle,\left[\nabla,\langle \vec{v}\rangle\right]\right] = \nabla\left(\overset{\downarrow}{\langle \vec{v}\rangle},\langle \vec{v}\rangle\right) - (\langle \vec{v}\rangle,\nabla)\overset{\downarrow}{\langle \vec{v}\rangle} = \frac{1}{2}\nabla|\langle \vec{v}\rangle|^2 - (\langle \vec{v}\rangle,\nabla)\langle \vec{v}\rangle,$$

$$(\langle \vec{v}\rangle,\nabla)\langle \vec{v}\rangle = \frac{1}{2}\nabla|\langle \vec{v}\rangle|^2 - \gamma\left[\langle \vec{v}\rangle,\text{rot}\,\vec{A}\right]. \qquad (51)$$

Considering (51), transform (49):

$$\nabla U = -\frac{1}{\beta}\left\{-\frac{1}{2\alpha}\frac{\partial \langle \vec{v} \rangle}{\partial t} - \frac{1}{4\alpha}\nabla|\langle \vec{v} \rangle|^2 + \alpha\nabla\left[\frac{\Delta\sqrt{f}}{\sqrt{f}}\right] + \frac{\gamma}{2\alpha}\frac{\partial \vec{A}}{\partial t} + \frac{1}{2\alpha}\nabla\frac{|\gamma \vec{A}|^2}{2}\right\} = \qquad (52)$$

$$= \frac{1}{2\alpha\beta}\left[\frac{\partial \langle \vec{v} \rangle}{\partial t} + (\langle \vec{v} \rangle, \nabla)\langle \vec{v} \rangle\right] + \frac{\gamma}{2\alpha\beta}\left[\langle \vec{v} \rangle, \text{rot}\, \vec{A}\right] - \frac{1}{2\alpha\beta}\nabla\frac{|\gamma \vec{A}|^2}{2} - \frac{\gamma}{2\alpha\beta}\frac{\partial \vec{A}}{\partial t} - \frac{\alpha}{\beta}\nabla\left[\frac{\Delta\sqrt{f}}{\sqrt{f}}\right].$$

It is important to consider that the total differential time from the velocity $\langle \vec{v} \rangle(\vec{r},t)$ has the form:

$$\frac{d}{dt}\langle \vec{v} \rangle(\vec{r},t) = \frac{\partial \langle \vec{v} \rangle}{\partial t} + (\langle \vec{v} \rangle, \nabla)\langle \vec{v} \rangle. \qquad (53)$$

Considering (53) and (43), expression (52) has the form:

$$-\frac{1}{2\alpha\beta}\frac{d\langle \vec{v} \rangle}{dt} = \nabla\left(-\frac{1}{2\alpha\beta}\frac{|\gamma\vec{A}|^2}{2} - U\right) - \frac{\gamma}{2\alpha\beta}\frac{\partial \vec{A}}{\partial t} + \frac{\gamma}{2\alpha\beta}\left[\langle \vec{v} \rangle, \text{rot}\, \vec{A}\right] - \frac{\alpha}{\beta}\nabla\left[\frac{\Delta\sqrt{f}}{\sqrt{f}}\right],$$

$$-\frac{1}{\gamma}\frac{d\langle \vec{v} \rangle}{dt} = -\frac{2\alpha\beta}{\gamma}\nabla\left(\frac{1}{2\alpha\beta}\frac{|\gamma\vec{A}|^2}{2} + U + \frac{\alpha}{\beta}\frac{\Delta\sqrt{f}}{\sqrt{f}}\right) - \frac{\partial \vec{A}}{\partial t} + \left[\langle \vec{v} \rangle, \text{rot}\, \vec{A}\right],$$

$$\frac{d\langle \vec{v} \rangle}{dt} = -\gamma\left(\vec{E} + \left[\langle \vec{v} \rangle, \vec{B}\right]\right), \qquad (54)$$

where

$$\vec{E} \stackrel{\text{det}}{=} -\nabla\chi - \frac{\partial \vec{A}}{\partial t},$$

$$\chi \stackrel{\text{det}}{=} \frac{2\alpha\beta}{\gamma}\left(\frac{1}{2\alpha\beta}\frac{|\gamma\vec{A}|^2}{2} + U + \frac{\alpha}{\beta}\left[\frac{\Delta\sqrt{f}}{\sqrt{f}}\right]\right). \qquad (55)$$

With all this, on account of notations (55) and (43), the following equations satisfy them:

$$\text{div}\, \vec{B} = 0, \qquad (56)$$

$$\text{rot}\, \vec{E} = -\frac{\partial}{\partial t}\text{rot}\, \vec{A} = -\frac{\partial \vec{B}}{\partial t} \qquad (57)$$

Transform the expression for $\chi$ function, considering (47):

$$\chi = \frac{2\alpha\beta}{\gamma}\left(\frac{1}{2\alpha\beta}\frac{|\gamma\vec{A}|^2}{2}+U+\frac{\alpha}{\beta}\left[\frac{\Delta\sqrt{f}}{\sqrt{f}}\right]\right)=\frac{2\alpha\beta}{\gamma}\left(\frac{1}{2\alpha\beta}\frac{|\gamma\vec{A}|^2}{2}-\frac{1}{\beta}\frac{\partial\varphi}{\partial t}+\frac{\alpha}{\beta}|\nabla\varphi|^2-\frac{\gamma}{\beta}(\vec{A},\nabla\varphi)\right)=$$

$$=\frac{2\alpha}{\gamma}\left(\frac{1}{2\alpha}\frac{|\gamma\vec{A}|^2}{2}-\frac{\partial\varphi}{\partial t}+\alpha|\nabla\varphi|^2-\gamma(\vec{A},\nabla\varphi)\right)=\frac{2\alpha}{\gamma}\left(\frac{1}{2\alpha}\frac{|\gamma\vec{A}|^2}{2}-\frac{\partial\varphi}{\partial t}+(\alpha\nabla\varphi,\nabla\varphi)-(\gamma\vec{A},\nabla\varphi)\right)=$$

$$=\frac{2\alpha}{\gamma}\left(\frac{1}{2\alpha}\frac{|\gamma\vec{A}|^2}{2}-\left(\frac{\partial\varphi}{\partial t}-(\alpha\nabla\varphi,\nabla\varphi)+(\gamma\vec{A},\nabla\varphi)\right)\right)=$$

$$=\frac{2\alpha}{\gamma}\left(\frac{1}{2\alpha}\frac{|\gamma\vec{A}|^2}{2}-\left(\frac{\partial\varphi}{\partial t}-(2\alpha\nabla\varphi,\nabla\varphi)+(\gamma\vec{A},\nabla\varphi)+\alpha(\nabla\varphi,\nabla\varphi)\right)\right)=$$

$$=\frac{2\alpha}{\gamma}\left(\frac{1}{2\alpha}\frac{|\gamma\vec{A}|^2}{2}-\left(\frac{\partial\varphi}{\partial t}+(-2\alpha\nabla\varphi+\gamma\vec{A},\nabla\varphi)+\alpha(\nabla\varphi,\nabla\varphi)\right)\right)=$$

$$=\frac{2\alpha}{\gamma}\left(\frac{1}{2\alpha}\frac{|\gamma\vec{A}|^2}{2}-\left(\frac{\partial\varphi}{\partial t}+(\langle\vec{v}\rangle,\nabla\varphi)+\alpha|\nabla\varphi|^2\right)\right),$$

i.e.

$$\chi = \frac{2\alpha}{\gamma}\left(\frac{1}{2\alpha}\frac{|\gamma\vec{A}|^2}{2}-\left(\frac{d\varphi}{dt}+\alpha|\nabla\varphi|^2\right)\right) \qquad (58)$$

It is necessary to calculate $\chi$ gradient to find $\vec{E}$ field. Now we intend to show the validity of the formula:

$$\nabla\frac{d}{dt}\varphi = \frac{d}{dt}\nabla\varphi - \alpha\nabla|\nabla\varphi|^2 + \nabla(\gamma\vec{A},\nabla\varphi) - (\gamma\vec{A},\nabla)\nabla\varphi. \qquad (59)$$

Indeed,

$$\frac{d}{dt}\nabla\varphi = \frac{d}{dt}\left(\frac{\partial\varphi}{\partial x}\vec{e}_x + \frac{\partial\varphi}{\partial y}\vec{e}_y + \frac{\partial\varphi}{\partial z}\vec{e}_z\right) =$$

$$= \vec{e}_x\left(\frac{\partial^2\varphi}{\partial x\partial t} + \frac{\partial^2\varphi}{\partial x^2}\frac{dx}{dt} + \frac{\partial^2\varphi}{\partial x\partial y}\frac{dy}{dt} + \frac{\partial^2\varphi}{\partial x\partial z}\frac{dz}{dt}\right) + \vec{e}_y\left(\frac{\partial^2\varphi}{\partial y\partial t} + \frac{\partial^2\varphi}{\partial y\partial x}\frac{dx}{dt} + \frac{\partial^2\varphi}{\partial y^2}\frac{dy}{dt} + \frac{\partial^2\varphi}{\partial y\partial z}\frac{dz}{dt}\right) +$$

$$+\vec{e}_z\left(\frac{\partial^2\varphi}{\partial z\partial t} + \frac{\partial^2\varphi}{\partial z\partial x}\frac{dx}{dt} + \frac{\partial^2\varphi}{\partial z\partial y}\frac{dy}{dt} + \frac{\partial^2\varphi}{\partial z^2}\frac{dz}{dt}\right) =$$

$$= \vec{e}_x\left(\frac{\partial^2\varphi}{\partial x\partial t} + \frac{\partial^2\varphi}{\partial x^2}\langle v_x\rangle + \frac{\partial^2\varphi}{\partial x\partial y}\langle v_y\rangle + \frac{\partial^2\varphi}{\partial x\partial z}\langle v_z\rangle\right) + \vec{e}_y\left(\frac{\partial^2\varphi}{\partial y\partial t} + \frac{\partial^2\varphi}{\partial y\partial x}\langle v_x\rangle + \frac{\partial^2\varphi}{\partial y^2}\langle v_y\rangle + \frac{\partial^2\varphi}{\partial y\partial z}\langle v_z\rangle\right) +$$

$$+\vec{e}_z\left(\frac{\partial^2\varphi}{\partial z\partial t} + \frac{\partial^2\varphi}{\partial z\partial x}\langle v_x\rangle + \frac{\partial^2\varphi}{\partial z\partial y}\langle v_y\rangle + \frac{\partial^2\varphi}{\partial z^2}\langle v_z\rangle\right). \qquad (60)$$

Write out the expression $\nabla \dfrac{d}{dt}\varphi$:

$$\nabla \frac{d}{dt}\varphi = \nabla\left(\frac{\partial \varphi}{\partial t} + (\langle \vec{v}\rangle, \nabla\varphi)\right) = \frac{\partial}{\partial t}\nabla\varphi + \nabla(\langle\vec{v}\rangle, \nabla\varphi) =$$

$$= \frac{\partial}{\partial t}\nabla\varphi + \nabla\left(\langle v_x\rangle\frac{\partial \varphi}{\partial x} + \langle v_y\rangle\frac{\partial \varphi}{\partial y} + \langle v_z\rangle\frac{\partial \varphi}{\partial z}\right) = \frac{\partial}{\partial t}\nabla\varphi +$$

$$+\vec{e}_x\left(\frac{\partial\langle v_x\rangle}{\partial x}\frac{\partial \varphi}{\partial x} + \langle v_x\rangle\frac{\partial^2 \varphi}{\partial x^2} + \frac{\partial\langle v_y\rangle}{\partial x}\frac{\partial \varphi}{\partial y} + \langle v_y\rangle\frac{\partial^2 \varphi}{\partial y\partial x} + \frac{\partial\langle v_z\rangle}{\partial x}\frac{\partial \varphi}{\partial z} + \langle v_z\rangle\frac{\partial^2 \varphi}{\partial z\partial x}\right)+ \qquad (61)$$

$$+\vec{e}_y\left(\frac{\partial\langle v_x\rangle}{\partial y}\frac{\partial \varphi}{\partial x} + \langle v_x\rangle\frac{\partial^2 \varphi}{\partial x\partial y} + \frac{\partial\langle v_y\rangle}{\partial y}\frac{\partial \varphi}{\partial y} + \langle v_y\rangle\frac{\partial^2 \varphi}{\partial y^2} + \frac{\partial\langle v_z\rangle}{\partial y}\frac{\partial \varphi}{\partial z} + \langle v_z\rangle\frac{\partial^2 \varphi}{\partial z\partial y}\right)+$$

$$+\vec{e}_z\left(\frac{\partial\langle v_x\rangle}{\partial z}\frac{\partial \varphi}{\partial x} + \langle v_x\rangle\frac{\partial^2 \varphi}{\partial x\partial z} + \frac{\partial\langle v_y\rangle}{\partial z}\frac{\partial \varphi}{\partial y} + \langle v_y\rangle\frac{\partial^2 \varphi}{\partial y\partial z} + \frac{\partial\langle v_z\rangle}{\partial z}\frac{\partial \varphi}{\partial z} + \langle v_z\rangle\frac{\partial^2 \varphi}{\partial z^2}\right)$$

Comparing (60) and (61) we obtain:

$$\nabla\frac{d}{dt}\varphi = \frac{d}{dt}\nabla\varphi + \vec{e}_x\left(\frac{\partial\langle v_x\rangle}{\partial x}\frac{\partial\varphi}{\partial x} + \frac{\partial\langle v_y\rangle}{\partial x}\frac{\partial\varphi}{\partial y} + \frac{\partial\langle v_z\rangle}{\partial x}\frac{\partial\varphi}{\partial z}\right)+$$

$$+\vec{e}_y\left(\frac{\partial\langle v_x\rangle}{\partial y}\frac{\partial\varphi}{\partial x} + \frac{\partial\langle v_y\rangle}{\partial y}\frac{\partial\varphi}{\partial y} + \frac{\partial\langle v_z\rangle}{\partial y}\frac{\partial\varphi}{\partial z}\right) + \vec{e}_z\left(\frac{\partial\langle v_x\rangle}{\partial z}\frac{\partial\varphi}{\partial x} + \frac{\partial\langle v_y\rangle}{\partial z}\frac{\partial\varphi}{\partial y} + \frac{\partial\langle v_z\rangle}{\partial z}\frac{\partial\varphi}{\partial z}\right) =$$

$$= \frac{d}{dt}\nabla\varphi + \vec{e}_x\frac{\partial}{\partial x}\left(\langle\overset{\downarrow}{v_x}\rangle\frac{\partial\varphi}{\partial x} + \langle\overset{\downarrow}{v_y}\rangle\frac{\partial\varphi}{\partial y} + \langle\overset{\downarrow}{v_z}\rangle\frac{\partial\varphi}{\partial z}\right)+$$

$$+\vec{e}_y\frac{\partial}{\partial y}\left(\langle\overset{\downarrow}{v_x}\rangle\frac{\partial\varphi}{\partial x} + \langle\overset{\downarrow}{v_y}\rangle\frac{\partial\varphi}{\partial y} + \langle\overset{\downarrow}{v_z}\rangle\frac{\partial\varphi}{\partial z}\right) + \vec{e}_z\frac{\partial}{\partial z}\left(\langle\overset{\downarrow}{v_x}\rangle\frac{\partial\varphi}{\partial x} + \langle\overset{\downarrow}{v_y}\rangle\frac{\partial\varphi}{\partial y} + \langle\overset{\downarrow}{v_z}\rangle\frac{\partial\varphi}{\partial z}\right) =$$

$$= \frac{d}{dt}\nabla\varphi + \nabla\left(\langle\overset{\downarrow}{\vec{v}}\rangle,\nabla\varphi\right) = \frac{d}{dt}\nabla\varphi + \nabla\left(-2\alpha\overset{\downarrow}{\nabla}\varphi + \gamma\vec{A},\nabla\varphi\right) = \frac{d}{dt}\nabla\varphi - 2\alpha\nabla\left(\overset{\downarrow}{\nabla}\varphi,\nabla\varphi\right)+ \qquad (62)$$

$$+\gamma\nabla\left(\overset{\downarrow}{\vec{A}},\nabla\varphi\right) = \frac{d}{dt}\nabla\varphi - \alpha\nabla|\nabla\varphi|^2 + \gamma\nabla\left(\overset{\downarrow}{\vec{A}},\nabla\varphi\right)$$

Then it is important to consider that

$$\nabla(\gamma\vec{A},\nabla\varphi) = \nabla\left(\gamma\vec{A},\overset{\downarrow}{\nabla}\varphi\right) + \nabla\left(\gamma\overset{\downarrow}{\vec{A}},\nabla\varphi\right) \Rightarrow$$

$$\nabla\left(\overset{\downarrow}{\vec{A}},\nabla\varphi\right) = \nabla(\gamma\vec{A},\nabla\varphi) - \nabla\left(\gamma\vec{A},\overset{\downarrow}{\nabla}\varphi\right) = \nabla(\gamma\vec{A},\nabla\varphi) - (\gamma\vec{A},\nabla)\nabla\varphi, \qquad (63)$$

as $\nabla\left(\gamma\vec{A},\vec{\nabla}\varphi\right)=\left(\gamma\vec{A},\nabla\right)\nabla\varphi$. Taking (63) into account, expression (62) transforms into (59), which required.

Using (59), the expression for $\nabla\chi$ takes the form:

$$\nabla\chi = \frac{2\alpha}{\gamma}\left(\frac{1}{2\alpha}\nabla\frac{|\gamma\vec{A}|^2}{2} - \left(\nabla\frac{d\varphi}{dt} + \alpha\nabla|\nabla\varphi|^2\right)\right) =$$

$$= \frac{2\alpha}{\gamma}\left(\frac{1}{2\alpha}\nabla\frac{|\gamma\vec{A}|^2}{2} - \left(\frac{d}{dt}\nabla\varphi - \alpha\nabla|\nabla\varphi|^2 + \nabla(\gamma\vec{A},\nabla\varphi) - (\gamma\vec{A},\nabla)\nabla\varphi + \alpha\nabla|\nabla\varphi|^2\right)\right) =$$

$$= \frac{2\alpha}{\gamma}\left(\frac{1}{2\alpha}\nabla\frac{|\gamma\vec{A}|^2}{2} - \frac{d}{dt}\nabla\varphi - \nabla(\gamma\vec{A},\nabla\varphi) + (\gamma\vec{A},\nabla)\nabla\varphi\right) \qquad (64)$$

So, the $\vec{E}$ field fully appears through the vortex component $\vec{A}$ and the potential $\nabla\varphi$ component of the velocity flow $\langle\vec{v}\rangle$, i.e.

$$\vec{E} = -\nabla\chi - \frac{\partial\vec{A}}{\partial t} =$$

$$= \frac{2\alpha}{\gamma}\left(-\frac{1}{2\alpha}\nabla\frac{|\gamma\vec{A}|^2}{2} + \frac{d}{dt}\nabla\varphi + \nabla(\gamma\vec{A},\nabla\varphi) - (\gamma\vec{A},\nabla)\nabla\varphi\right) - \frac{\partial\vec{A}}{\partial t} \qquad (65)$$

If we introduce the notations

$$\langle\vec{v}_p\rangle = -2\alpha\nabla\varphi,\ \langle\vec{v}_s\rangle = \gamma\vec{A},$$
$$\langle\vec{v}\rangle = \langle\vec{v}_p\rangle + \langle\vec{v}_s\rangle, \qquad (66)$$

then the expression for the $\vec{E}$ field will be:

$$\vec{E} = \frac{2\alpha}{\gamma}\left(-\frac{1}{2\alpha}\nabla\frac{|\langle\vec{v}_s\rangle|^2}{2} - \frac{1}{2\alpha}\frac{d}{dt}\langle\vec{v}_p\rangle - \frac{1}{2\alpha}\nabla(\langle\vec{v}_s\rangle,\langle\vec{v}_p\rangle) + \frac{1}{2\alpha}(\langle\vec{v}_s\rangle,\nabla)\langle\vec{v}_p\rangle\right) - \frac{1}{\gamma}\frac{\partial\langle\vec{v}_s\rangle}{\partial t} =$$

$$= \frac{1}{\gamma}\left(-\nabla\frac{|\langle\vec{v}_s\rangle|^2}{2} - \frac{d}{dt}\langle\vec{v}_p\rangle - \nabla(\langle\vec{v}_s\rangle,\langle\vec{v}_p\rangle) + (\langle\vec{v}_s\rangle,\nabla)\langle\vec{v}_p\rangle - \frac{\partial\langle\vec{v}_s\rangle}{\partial t}\right) =$$

$$= \frac{1}{\gamma}\left(-\nabla\frac{|\langle\vec{v}_s\rangle|^2}{2} - \frac{\partial(\langle\vec{v}_p\rangle + \langle\vec{v}_s\rangle)}{\partial t} - (\langle\vec{v}\rangle,\nabla)\langle\vec{v}_p\rangle - \nabla(\langle\vec{v}_s\rangle,\langle\vec{v}_p\rangle) + (\langle\vec{v}_s\rangle,\nabla)\langle\vec{v}_p\rangle\right) =$$

$$= \frac{1}{\gamma}\left(-\nabla\frac{|\langle\vec{v}_s\rangle|^2}{2} - \frac{\partial\langle\vec{v}\rangle}{\partial t} - (\langle\vec{v}_s\rangle,\nabla)\langle\vec{v}_p\rangle - (\langle\vec{v}_p\rangle,\nabla)\langle\vec{v}_p\rangle - \nabla(\langle\vec{v}_s\rangle,\langle\vec{v}_p\rangle) + (\langle\vec{v}_s\rangle,\nabla)\langle\vec{v}_p\rangle\right) =$$

$$= -\frac{1}{\gamma}\left(\nabla \frac{|\langle \vec{v}_s \rangle|^2}{2} + \frac{\partial \langle \vec{v} \rangle}{\partial t} + \left(\langle \vec{v}_p \rangle, \nabla\right)\langle \vec{v}_p \rangle + \nabla\left(\langle \vec{v}_s \rangle, \langle \vec{v}_p \rangle\right)\right) \tag{67}$$

Using (51) for $\langle \vec{v}_p \rangle$, there will be

$$\left(\langle \vec{v}_p \rangle, \nabla\right)\langle \vec{v}_p \rangle = \frac{1}{2}\nabla |\langle \vec{v}_p \rangle|^2 \tag{68}$$

Considering (68), expression (67) takes the form

$$\vec{E} = -\frac{1}{\gamma}\left(\frac{1}{2}\nabla\left(|\langle \vec{v}_s \rangle|^2 + |\langle \vec{v}_p \rangle|^2\right) + \frac{\partial \langle \vec{v} \rangle}{\partial t} + \nabla\left(\langle \vec{v}_s \rangle, \langle \vec{v}_p \rangle\right)\right) =$$

$$= -\frac{1}{\gamma}\left(\frac{\partial \langle \vec{v} \rangle}{\partial t} + \frac{1}{2}\nabla\left(|\langle \vec{v}_s \rangle|^2 + 2\left(\langle \vec{v}_s \rangle, \langle \vec{v}_p \rangle\right) + |\langle \vec{v}_p \rangle|^2\right)\right) =$$

$$= -\frac{1}{\gamma}\left(\frac{\partial \langle \vec{v} \rangle}{\partial t} + \frac{1}{2}\nabla\left(\langle \vec{v}_s \rangle + \langle \vec{v}_p \rangle, \langle \vec{v}_s \rangle + \langle \vec{v}_p \rangle\right)\right),$$

i.e.

$$\vec{E} = -\frac{1}{\gamma}\left(\frac{\partial \langle \vec{v} \rangle}{\partial t} + \nabla \frac{|\langle \vec{v} \rangle|^2}{2}\right) \tag{69}$$

(expression (69) is completely similar to the expression from [7]).
For the $\chi$ function, apart from (55) and (58), it is possible to obtain the expression with the velocity $\langle \vec{v}_s \rangle$ and $\langle \vec{v}_p \rangle$. Using (66):

$$\vec{E} = -\frac{1}{\gamma}\nabla\frac{|\langle \vec{v} \rangle|^2}{2} - \frac{1}{\gamma}\frac{\partial \langle \vec{v}_p \rangle}{\partial t} - \frac{1}{\gamma}\frac{\partial \langle \vec{v}_s \rangle}{\partial t} = -\frac{1}{\gamma}\nabla\left(\frac{|\langle \vec{v} \rangle|^2}{2} - 2\alpha\frac{\partial \varphi}{\partial t}\right) - \frac{\partial \vec{A}}{\partial t}. \tag{70}$$

Comparing (70) and (55), we obtain the following

$$\chi = \frac{1}{\gamma}\left(\frac{|\langle \vec{v} \rangle|^2}{2} - 2\alpha\frac{\partial \varphi}{\partial t}\right) \tag{71}$$

If the field of velocities does not have the vortex component $\vec{A}$, then (65) is real

$$\nabla \chi = -\frac{2\alpha}{\gamma}\frac{d}{dt}\nabla \varphi = \frac{1}{\gamma}\frac{d}{dt}(-2\alpha \nabla \varphi) = \frac{1}{\gamma}\frac{d\langle \vec{v} \rangle}{dt}$$

or

$$\frac{d\langle \vec{v} \rangle}{dt} = \gamma \nabla \chi$$

that corresponds to the irrotational field equation $\langle \vec{v} \rangle$.

## 4. The ideology of the self-consistent problems

***Definition.*** Let function $f(\vec{r},t)$ and vector-function $\langle \vec{v} \rangle(\vec{r},t)$ satisfy equation (1). We will call them strongly-agreed functions, provided the conditions are realized:

$$\frac{\partial \vec{D}}{\partial t} + f \langle \vec{v} \rangle = \text{rot}\, \vec{H}, \tag{72}$$

$$\text{div}\, \vec{D}(\vec{r},t) = f(\vec{r},t) \tag{73}$$

$$\vec{D}(\vec{r},t) = \bar{\varepsilon} \vec{E}(\vec{r},t),\ \vec{B}(\vec{r},t) = \text{rot}\, \vec{A}(\vec{r},t) = \bar{\mu} \vec{H}(\vec{r},t),$$

where $\bar{\varepsilon}$ and $\bar{\mu}$ are certain constant values. If not, we will call them weakly-agreed functions.

Note that at taking the divergence from equation (72) and account of (73), it transforms into equation (1). In the general way, the $\vec{D}(\vec{r},t)$ field divergence is different from $f(\vec{r},t)$ and has the form:

$$\text{div}\, \vec{D} = \bar{\varepsilon}\, \text{div}\, \vec{E} = -\frac{\bar{\varepsilon}}{\gamma} \left( \frac{1}{2} \Delta |\langle \vec{v} \rangle|^2 + \frac{\partial}{\partial t} \text{div}\, \langle \vec{v} \rangle \right) = -\frac{\bar{\varepsilon}}{\gamma} \left( \frac{1}{2} \Delta |\langle \vec{v} \rangle|^2 - 2\alpha \frac{\partial}{\partial t} \Delta \varphi \right) =$$

$$= -\frac{\bar{\varepsilon}}{\gamma} \Delta \left( \frac{1}{2} |\langle \vec{v} \rangle|^2 - 2\alpha \frac{\partial \varphi}{\partial t} \right) = -\bar{\varepsilon} \Delta \chi \tag{74}$$

The examples of strongly-agreed and weakly-agreed functions $f(\vec{r},t)$ and $\langle \vec{v} \rangle(\vec{r},t)$ are given below.

**Example 1.** (weakly-agreed functions)
Let the vector-function $\langle \vec{v} \rangle(\vec{r},t)$ has the form:

$$\langle \vec{v} \rangle(\vec{r},t) = at \vec{e}_x, \tag{75}$$

where $a$ − constant value. From (75) it is seen that the a function $f$ as an equation (1) solving can be derived by the method of characteristics. Indeed, equation (1) with the account of (75) takes the form:

$$f_t + at f_x = 0 \tag{76}$$

Therefore

$$\frac{dx}{dt} = at,\ \xi(x,t) = x - x_0 - \frac{at^2}{2} \tag{77}$$

The equations $\xi(x,t) = Const$ set the characteristics, along which $f(x,t)$ will be constant. Indeed:

$$f(x,t) = F(\xi(x,t)),$$

$$F'\xi_t + atF'\xi_x = F'(\xi_t + at\xi_x) = F'(-at + at) = 0,$$

at a random, continuously differentiable function $F(\xi)$. To fix the idea, as $F(\xi)$ we will take

$$F(\xi) = e^{-\xi^2},$$

$$f(x,t) = e^{-\left(x-x_0-\frac{at^2}{2}\right)^2}. \tag{78}$$

Functions (75) and (78) satisfy equation (1). From (75), (4) and (7) it is seen that

$$\varphi(x,t) = -\frac{atx}{2\alpha} + const, \quad \vec{A}(x,t) = \vec{0} \tag{79}$$

Substituting (78), (79) into (47) there will be an expression for $U$ function

$$U(x,t) = -\frac{1}{\beta}\left\{\frac{\partial\varphi}{\partial t} + \alpha\frac{\Delta\sqrt{f}}{\sqrt{f}} - \alpha|\nabla\varphi(\vec{r},t)|^2\right\} = -\frac{1}{\beta}\left\{-\frac{ax}{2\alpha} + \alpha e^{\frac{\xi^2}{2}}\Delta e^{-\frac{\xi^2}{2}} - \alpha\frac{a^2t^2}{4\alpha^2}\right\} =$$

$$= \frac{ax}{2\alpha\beta} + \frac{a^2t^2}{4\alpha\beta} + \frac{\alpha}{\beta}e^{\frac{\xi^2}{2}}\mathrm{div}\left(e^{-\frac{\xi^2}{2}}\xi\xi_x\vec{e}_x\right) = \frac{ax}{2\alpha\beta} + \frac{a^2t^2}{4\alpha\beta} + \frac{\alpha}{\beta}e^{\frac{\xi^2}{2}}e^{-\frac{\xi^2}{2}}\left(\xi_x^2 - \xi^2\xi_x^2 + \xi\xi_{xx}\right),$$

$$U(x,t) = \frac{ax}{2\alpha\beta} + \frac{a^2t^2}{4\alpha\beta} + \frac{\alpha}{\beta}\left(1 - \xi(x,t)^2\right). \tag{80}$$

The function $\Psi(x,t)$ according to (3), (6), (78), and (79) has the form:

$$\Psi(x,t) = C_0 e^{-\frac{\xi(x,t)^2}{2} - i\frac{atx}{2\alpha}}, \tag{81}$$

where $C_0 = const$. Thus, the function (81) satisfies equation (28), which has the following form here:

$$\frac{i}{\beta}\frac{\partial\Psi}{\partial t} = \frac{\alpha}{\beta}\Delta\Psi + U\Psi. \tag{82}$$

Let us demonstrate that functions (75) and (78) are weakly-agreed. Check the condition (73) finding $\vec{E}$, e.g. by formula (69)

$$\vec{E} = -\frac{1}{\gamma}\left(\frac{\partial\langle\vec{v}\rangle}{\partial t} + \nabla\frac{|\langle\vec{v}\rangle|^2}{2}\right) = -\frac{a}{\gamma}\vec{e}_x, \tag{83}$$

or the same by formula (55)

$$\vec{E} = -\frac{2\alpha\beta}{\gamma}\nabla\left(U + \frac{\alpha}{\beta}\left[\frac{\Delta\sqrt{f}}{\sqrt{f}}\right]\right) = -\frac{2\alpha\beta}{\gamma}\nabla\left(\frac{ax}{2\alpha\beta} + \frac{a^2t^2}{4\alpha\beta}\right) = -\frac{a}{\gamma}\vec{e}_x. \tag{84}$$

In both (83) and (84) variants there is the same result. Check the condition (73) and obtain:

$$\mathrm{div}\,\vec{D}(\vec{r},t) = \bar{\varepsilon}\,\mathrm{div}\,\vec{E}(\vec{r},t) = -\bar{\varepsilon}\frac{a}{\gamma}\,\mathrm{div}\,\vec{e}_x = 0 \neq f(x,t).$$

In that way, the condition (73) is not proved. Therefore, functions (75) and (78) are weakly-agreed.

**Example 2**. (strongly-agreed functions)

Regard the example from [7, 8] evolution of the density function of the homogeneously charged sphere charge. Take $\langle \vec{v} \rangle (r,t)$ as a vector-function:

$$\langle \vec{v} \rangle (r,t) = b(t)\,r\vec{e}_r, \qquad (85)$$

where the function $b(t)$ is to be determined. From (85) it is seen that

$$\varphi = -\frac{b(t)}{4\alpha}r^2 + Const, \quad \vec{A}(r,t) = \vec{0}. \qquad (86)$$

As $f$ function we take:

$$f(r,t) = 3a(t), \qquad (87)$$

where the function $a(t)$ is to be determined. Substituting (85) and (87) into equation (1) we obtain:

$$3\frac{\partial a}{\partial t} + 3ab\,\mathrm{div}\,r\vec{e}_r = 0,$$

or

$$a' + 3ab = 0. \qquad (88)$$

According to (88), function $a(t)$ can be presented in the form of:

$$a(t) = C_0 \exp\left(-3\int_{t_0}^{t} b\,dt\right). \qquad (89)$$

In this way, knowing the function $b(t)$, we can find the function $a(t)$ and vice versa.

As $\vec{D}$ field we will take

$$\vec{D}(r,t) = a(t)\,r\vec{e}_r. \qquad (90)$$

Substituting (90) into (72) we obtain equation (88), i.e. true identity. Substituting (90) into (73) owing to (87), we also obtain the true identity. Therefore, functions (85) and (87) are strongly-agreed. To determine function $a(t)$ we use the equation (54):

$$\frac{\partial \langle \vec{v} \rangle}{\partial t} + (\langle \vec{v} \rangle, \nabla)\langle \vec{v} \rangle = (b' + b^2)r\vec{e}_r = -\frac{\gamma}{\varepsilon}\vec{D} = -\frac{\gamma}{\varepsilon}a(t)r\vec{e}_r,$$

$$(b' + b^2) = -\frac{\gamma}{\varepsilon}a(t). \tag{91}$$

Expressing $b(t)$ from (88) and substituting into (91), we obtain the equation relative to $a(t)$:

$$3aa_{tt} - 4a_t^2 - 9\frac{\gamma}{\varepsilon}a^3 = 0 \tag{92}$$

Interchange equation (92)

$$f(r,t) = 3a(t) = f_0\left(\frac{R_0}{R(t)}\right)^3 = \frac{Q}{4\pi}\frac{1}{R(t)^3}, \tag{93}$$

where $Q$ – total charge of the sphere with the initial radius $R_0$ and $f_0 = f(r,0) = \frac{3Q}{4\pi R_0^3}$ – initial density of the charge in the sphere. Interchanging (93), equation (93) transforms into [7, 8]:

$$R_{tt} = \bar{\gamma}\frac{1}{R^2}, \tag{94}$$

where $\bar{\gamma} = -\frac{\gamma Q}{4\pi\bar{\varepsilon}}$. Equation (94) has a solving in the form of [7, 8]

$$t = \frac{R_0^{3/2}}{\sqrt{2\bar{\gamma}}}\left(\sqrt{\frac{R(t)}{R_0}\left(\frac{R(t)}{R_0} - 1\right)} + \operatorname{arcch}\sqrt{\frac{R(t)}{R_0}}\right), \tag{95}$$

or

$$t = \frac{R_0^{3/2}}{\sqrt{2\bar{\gamma}}}\left(\sqrt{\left(\frac{f}{f_0}\right)^{1/3}\left(\left(\frac{f}{f_0}\right)^{1/3} - 1\right)} + \operatorname{arcch}\left(\frac{f}{f_0}\right)^{1/6}\right).$$

Note that $f$ inside the sphere is a constant value and depends only on time (compare with (93)). The function $\Psi(r,t)$ has the form:

$$\Psi(\vec{r},t) = \sqrt{3a(t)}e^{-i\frac{b(t)}{4\alpha}r^2}, \tag{96}$$

and satisfies the equation (82) where $U$ has the form:

$$U(r,t) = -\frac{1}{\beta}\left\{-\frac{r^2}{4\alpha}b' + \alpha\left[\frac{\Delta\sqrt{3a}}{\sqrt{3a}} - \frac{b^2 r^2}{4\alpha^2}\right]\right\} = \frac{r^2}{4\alpha\beta}(b' + b^2). \tag{97}$$

## 5. The equation of the center of mass movement

Average the equation (54) on the spatial coordinate.

$$\int f(\vec{r},t)\frac{d\langle\vec{v}\rangle}{dt}(\vec{r},t)d^3r = -\gamma\int f(\vec{r},t)\vec{E}(\vec{r},t)d^3r - \gamma\int f(\vec{r},t)\left[\langle\vec{v}\rangle(\vec{r},t),\vec{B}(\vec{r},t)\right]d^3r. \qquad (98)$$

Simplify the left part of equation (98). Considering equation (1), write down integrand function:

$$f(\vec{r},t)\frac{d\langle\vec{v}\rangle}{dt}(\vec{r},t) = f\frac{\partial\langle\vec{v}\rangle}{\partial t} + f(\langle\vec{v}\rangle,\nabla)\langle\vec{v}\rangle = \frac{\partial f\langle\vec{v}\rangle}{\partial t} - \langle\vec{v}\rangle\frac{\partial f}{\partial t} + f(\langle\vec{v}\rangle,\nabla)\langle\vec{v}\rangle =$$

$$= \frac{\partial f\langle\vec{v}\rangle}{\partial t} + \langle\vec{v}\rangle\operatorname{div}\left[f\langle\vec{v}\rangle\right] + f(\langle\vec{v}\rangle,\nabla)\langle\vec{v}\rangle. \qquad (99)$$

Demonstrate that the following correlation is true:

$$\int\langle\vec{v}\rangle\operatorname{div}_r\left[\langle\vec{v}\rangle f\right]d^3r = -\int f(\langle\vec{v}\rangle,\nabla)\langle\vec{v}\rangle d^3r. \qquad (100)$$

Check (100) by the direct integration.

$$\int\langle\vec{v}\rangle\operatorname{div}_r\left[\langle\vec{v}\rangle f\right]d^3r = \vec{e}_x\int\langle v_x\rangle\left[\frac{\partial}{\partial x}(f\langle v_x\rangle) + \frac{\partial}{\partial y}(f\langle v_y\rangle) + \frac{\partial}{\partial z}(f\langle v_z\rangle)\right]d^3r +$$

$$+ \vec{e}_y\int\langle v_y\rangle\left[\frac{\partial}{\partial x}(f\langle v_x\rangle) + \frac{\partial}{\partial y}(f\langle v_y\rangle) + \frac{\partial}{\partial z}(f\langle v_z\rangle)\right]d^3r + \qquad (101)$$

$$+ \vec{e}_z\int\langle v_z\rangle\left[\frac{\partial}{\partial x}(f\langle v_x\rangle) + \frac{\partial}{\partial y}(f\langle v_y\rangle) + \frac{\partial}{\partial z}(f\langle v_z\rangle)\right]d^3r.$$

Write down (101) the integral only for $\vec{e}_x$ component, the rest will be transformed in the similar way.

$$\int\langle v_x\rangle\left[\frac{\partial}{\partial x}(f\langle v_x\rangle) + \frac{\partial}{\partial y}(f\langle v_y\rangle) + \frac{\partial}{\partial z}(f\langle v_z\rangle)\right]d^3r =$$

$$= \int\langle v_x\rangle\frac{\partial}{\partial x}(f\langle v_x\rangle)dx\int dy\int dz + \int dx\int\langle v_x\rangle\frac{\partial}{\partial y}(f\langle v_y\rangle)dy\int dz + \int dx\int dy\int\langle v_x\rangle\frac{\partial}{\partial z}(f\langle v_z\rangle)dz =$$

$$= \int dy\int dz\left[\langle v_x\rangle^2 f\Big|_{-\infty}^{+\infty} - \int f\langle v_x\rangle\frac{\partial\langle v_x\rangle}{\partial x}dx\right] + \int dx\int dz\left[\langle v_x\rangle\langle v_y\rangle f\Big|_{-\infty}^{+\infty} - \int f\langle v_y\rangle\frac{\partial\langle v_x\rangle}{\partial y}dy\right] +$$

$$+ \int dx\int dy\left[\langle v_x\rangle\langle v_z\rangle f\Big|_{-\infty}^{+\infty} - \int f\langle v_z\rangle\frac{\partial\langle v_x\rangle}{\partial z}dz\right] =$$

$$= -\int f\langle v_x\rangle\frac{\partial\langle v_x\rangle}{\partial x}d^3r - \int f\langle v_y\rangle\frac{\partial\langle v_x\rangle}{\partial y}d^3r - \int f\langle v_z\rangle\frac{\partial\langle v_x\rangle}{\partial z}d^3r = -\int f(\langle\vec{v}\rangle,\nabla)\langle v_x\rangle d^3r. \qquad (102)$$

The same expressions (102) will be for the components $\vec{e}_y$ and $\vec{e}_z$, as a result (101) will be:

$$\int \langle \vec{v} \rangle \mathrm{div}_r \left[ \langle \vec{v} \rangle f \right] d^3r = -\int f(\langle \vec{v} \rangle, \nabla) \langle v_x \rangle \vec{e}_x d^3r - \int f(\langle \vec{v} \rangle, \nabla) \langle v_y \rangle \vec{e}_y d^3r -$$
$$- \int f(\langle \vec{v} \rangle, \nabla) \langle v_z \rangle \vec{e}_z d^3r = -\int f(\langle \vec{v} \rangle, \nabla) \langle \vec{v} \rangle d^3r, \tag{103}$$

which required. In the paper [7] was seen that:

$$\frac{d}{dt}\left[ N(t) \langle\langle \vec{v} \rangle\rangle(t) \right] = N(t) \langle\langle\langle \dot{\vec{v}} \rangle\rangle\rangle(t), \tag{104}$$

where

$$N(t) = \int f(\vec{r},t) d^3r = \int \bar{\Psi}(\vec{r},t) \Psi(\vec{r},t) d^3r,$$

$$\langle \dot{\vec{v}} \rangle(\vec{r},\vec{v},t) = \frac{1}{f_2(\vec{r},\vec{v},t)} \int \dot{\vec{v}} f_3(\vec{r},\vec{v},\dot{\vec{v}},t) d^3\dot{v},$$

$$\langle\langle \dot{\vec{v}} \rangle\rangle(\vec{r},t) = \frac{1}{f_1(\vec{r},t)} \int \dot{\vec{v}} f_3(\vec{r},\vec{v},\dot{\vec{v}},t) d^3v d^3\dot{v} = \frac{1}{f_1(\vec{r},t)} \int \langle \dot{\vec{v}} \rangle(\vec{r},\vec{v},t) f_2(\vec{r},\vec{v},t) d^3v,$$

$$\langle\langle\langle \dot{\vec{v}} \rangle\rangle\rangle(t) = \frac{1}{N(t)} \int \dot{\vec{v}} f_3(\vec{r},\vec{v},\dot{\vec{v}},t) d^3r d^3v d^3\dot{v},$$

under condition that functions $f_1(\vec{r},t)$, $f_2(\vec{r},\vec{v},t)$ and $f_3(\vec{r},\vec{v},\dot{\vec{v}},t)$ satisfy the corresponding equations at the Vlasov chain of equations.

In this way, the left part of equation (98) considering (99), (100), and (104) transform into

$$\int f(\vec{r},t) \frac{d\langle \vec{v} \rangle}{dt}(\vec{r},t) d^3r = \frac{\partial}{\partial t} \int f \langle \vec{v} \rangle d^3r + \int \langle \vec{v} \rangle \mathrm{div}\left[ f \langle \vec{v} \rangle \right] d^3r + \int f(\langle \vec{v} \rangle, \nabla) \langle \vec{v} \rangle d^3r =$$
$$= \frac{d}{dt}\left[ N(t) \langle\langle \vec{v} \rangle\rangle(t) \right] - \int f(\langle \vec{v} \rangle, \nabla) \langle \vec{v} \rangle d^3r + \int f(\langle \vec{v} \rangle, \nabla) \langle \vec{v} \rangle d^3r = N(t) \langle\langle\langle \dot{\vec{v}} \rangle\rangle\rangle(t). \tag{105}$$

Transform the right part of equation (98):

$$-\gamma \int f(\vec{r},t) \vec{E}(\vec{r},t) d^3r - \gamma \int f(\vec{r},t) \left[ \langle \vec{v} \rangle(\vec{r},t), \vec{B}(\vec{r},t) \right] d^3r =$$
$$= -\gamma N(t) \left( \langle \vec{E} \rangle(t) + \langle [\langle \vec{v} \rangle, \vec{B}] \rangle(t) \right). \tag{106}$$

According to this, equation (98) has the form:

$$\langle\langle\langle \dot{\vec{v}} \rangle\rangle\rangle(t) = -\gamma \left( \langle \vec{E} \rangle(t) + \langle [\langle \vec{v} \rangle, \vec{B}] \rangle(t) \right). \tag{107}$$

The expression $\langle \vec{E} \rangle(t)$ can be simplified. According to (55) we obtain

$$\int f(\vec{r},t) \vec{E}(\vec{r},t) d^3r = \int f \left\{ -\nabla \frac{1}{\gamma} \left[ \frac{|\gamma \vec{A}|^2}{2} + 2\alpha\beta U + 2\alpha^2 \frac{\Delta|\Psi|}{|\Psi|} \right] - \frac{\partial \vec{A}}{\partial t} \right\} d^3r =$$

$$= \int f \left\{ -\nabla \frac{1}{\gamma} \left[ \frac{|\gamma \vec{A}|^2}{2} + 2\alpha\beta U \right] - \frac{\partial \vec{A}}{\partial t} \right\} d^3r - \frac{2\alpha^2}{\gamma} \int f \nabla \left( \frac{\Delta |\Psi|}{|\Psi|} \right) d^3r. \tag{108}$$

Let us demonstrate that the second integral at (108) equal to zero.

$$\int f \nabla \left[ \frac{\Delta |\Psi|}{|\Psi|} \right] d^3r = \int |\Psi|^2 \nabla \left[ \frac{\Delta |\Psi|}{|\Psi|} \right] d^3r =$$

$$= \vec{e}_x \int |\Psi|^2 \frac{\partial}{\partial x} \left[ \frac{\Delta |\Psi|}{|\Psi|} \right] dx \int dy \int dz + \vec{e}_y \int dx \int |\Psi|^2 \frac{\partial}{\partial y} \left[ \frac{\Delta |\Psi|}{|\Psi|} \right] dy \int dz +$$

$$+ \vec{e}_z \int dx \int dy \int |\Psi|^2 \frac{\partial}{\partial z} \left[ \frac{\Delta |\Psi|}{|\Psi|} \right] dz =$$

$$= \vec{e}_x \int dy \int dz \left[ |\Psi| \Delta |\Psi| \Big|_{-\infty}^{+\infty} - 2 \int \Delta |\Psi| \frac{\partial |\Psi|}{\partial x} dx \right] + \vec{e}_y \int dx \int dz \left[ |\Psi| \Delta |\Psi| \Big|_{-\infty}^{+\infty} - 2 \int \Delta |\Psi| \frac{\partial |\Psi|}{\partial y} dy \right] +$$

$$+ \vec{e}_z \int dx \int dy \left[ |\Psi| \Delta |\Psi| \Big|_{-\infty}^{+\infty} - 2 \int \Delta |\Psi| \frac{\partial |\Psi|}{\partial z} dz \right] =$$

$$= -2\vec{e}_x \int dy \int dz \int \Delta |\Psi| \frac{\partial |\Psi|}{\partial x} dx - 2\vec{e}_y \int dx \int dz \int \Delta |\Psi| \frac{\partial |\Psi|}{\partial y} dy - 2\vec{e}_z \int dx \int dy \int \Delta |\Psi| \frac{\partial |\Psi|}{\partial z} dz =$$

$$= -2 \int \Delta |\Psi| \nabla |\Psi| d^3r.$$

In this way, we have:

$$\int f \nabla \left[ \frac{\Delta |\Psi|}{|\Psi|} \right] d^3r = -2 \int \Delta |\Psi| \nabla |\Psi| d^3r. \tag{109}$$

For the convenient transformation of (109) introduce the notation $w \stackrel{det}{=} |\Psi|$, $\vec{Y} \stackrel{det}{=} -\nabla w$, $\sigma \stackrel{det}{=} \operatorname{div} \vec{Y} = -\Delta w$ then

$$\int \Delta |\Psi| \nabla |\Psi| d^3r = \int \sigma \vec{Y} d^3r = \int \vec{Y} \operatorname{div} \vec{Y} d^3r. \tag{110}$$

Show that the integral (110) equal to zero. Indeed, according to the vector $\vec{Y}$ it is seen that $\operatorname{rot} \vec{Y} = 0$. Then the formula (100) at $f \equiv 1$ is to satisfy the integral (110). Additionally take (51) in the formula (100):

$$(\vec{Y}, \nabla) \vec{Y} = \frac{1}{2} \nabla |\vec{Y}|^2 - [\vec{Y}, \operatorname{rot} \vec{Y}] = \frac{1}{2} \nabla |\vec{Y}|^2,$$

$$\int \vec{Y} \operatorname{div} \vec{Y} d^3r = -\frac{1}{2} \int \nabla Y^2 d^3r = -\frac{1}{2} \vec{e}_x \int \frac{\partial Y^2}{\partial x} dx \int dy \int dz - \frac{1}{2} \vec{e}_y \int dx \int \frac{\partial Y^2}{\partial y} dy \int dz -$$

$$-\frac{1}{2}\vec{e}_z \int dx \int dy \int \frac{\partial Y^2}{\partial z} dz = -\frac{1}{2}\vec{e}_x \int dy \int dz \left[Y^2 \Big|_{-\infty}^{+\infty}\right] - \frac{1}{2}\vec{e}_y \int dx \int dz \left[Y^2 \Big|_{-\infty}^{+\infty}\right] -$$
$$-\frac{1}{2}\vec{e}_z \int dx \int dy \left[Y^2 \Big|_{-\infty}^{+\infty}\right] = 0, \qquad (111)$$

which required. As a result, expression (108) takes the form

$$\int f(\vec{r},t)\vec{E}(\vec{r},t)d^3r = \int f \left\{ -\nabla \frac{1}{\gamma} \left[ \frac{|\gamma \vec{A}|^2}{2} + 2\alpha\beta U \right] - \frac{\partial \vec{A}}{\partial t} \right\} d^3r. \qquad (112)$$

In case of irrotational field, the equation (107) will be

$$-\frac{1}{2\alpha\beta} \langle\langle\langle \dot{\vec{v}} \rangle\rangle\rangle (t) = -\langle \nabla U \rangle (t). \qquad (113)$$

### 6. Special cases

Take a look at special cases of equations (29) and (41). Denote them

$$\alpha = -\frac{\hbar}{2m}, \quad \beta = \frac{1}{\hbar} \text{ и } \gamma = -\frac{e}{m}. \qquad (114)$$

If the vector field $\langle \vec{v} \rangle$ has only a potential part at expansion (4), then there is $\vec{A} = \vec{\Theta}$, and equation (29) has the form:

$$i\hbar \frac{\partial \Psi}{\partial t} = \left[ \frac{\hat{p}^2}{2m} + U \right] \Psi = \hat{H}\Psi. \qquad (115)$$

The equation (43) is congruent with the well-known Schrödinger equation for the wave function $\Psi$. If there is a vortex (solenoidal) velocity component at expansion (4), we obtain the equation (29) with $T_{rot} = \frac{mv_s^2}{2}$ value. From the physical point of view, $T_{rot}$ value corresponds with the kinematic energy of the vortex (in special case rotary) movement.

$$i\hbar \frac{\partial \Psi}{\partial t} = \left[ \frac{1}{2m}\left(\hat{p} - \frac{e}{c}\vec{A}\right)^2 - \frac{mv_s^2}{2} + U \right] \Psi = \hat{H}\Psi, \qquad (116)$$

or

$$i\hbar \frac{\partial \Psi}{\partial t} = \left[ \frac{1}{2m}\left(\hat{p} - \frac{e}{c}\vec{A}\right)^2 - L \right] \Psi = \hat{H}\Psi,$$

where $L = T_{rot} - U$. Take a look at a special case of the rotation of the particle with the charge $e$ and mass $m$ in the constant magnetic field $B_0$. The following correlations are true here:

$$m\frac{v_s^2}{r} = ev_s B_0,$$

where $r$ – circle radius, which is equal to the introduced Compton wavelength, i.e. $r = \bar{\lambda}_C = \frac{\hbar}{mc}$. The velocity $v_s$ is formally equal to the velocity of light $c$, then

$$\frac{1}{2\alpha\beta}\frac{\left|\gamma\vec{A}\right|^2}{2} = -\frac{mv_s^2}{2} = -\frac{ev_s B_0 \bar{\lambda}_C}{2} = -\frac{ev_s B_0 \hbar}{2mc} = -\frac{e\hbar}{2m}B_0 \qquad (117)$$

Substituting (117) into equation (116), we get

$$i\hbar\frac{\partial\Psi}{\partial t} = \left[\frac{1}{2m}\left(\hat{p} - \frac{e}{c}\vec{A}\right)^2 - \frac{e\hbar}{2m}B_0 + U\right]\Psi = \hat{H}\Psi, \qquad (118)$$

In formal terms, equation (118) coincides with the equation for one of the spinar components in the Pauli equation, describing the particle with the spin.

Mean velocity (4) considering the notations for $\alpha, \gamma$ and velocity potential $\Phi$ (7), will be

$$\langle\vec{v}\rangle(\vec{r},t) = \frac{\hbar}{2m}\nabla\Phi(\vec{r},t) - \frac{e}{m}\vec{A}(\vec{r},t) = \frac{\hbar}{m}\nabla\varphi(r,t) - \frac{e}{m}\vec{A}(\vec{r},t), \qquad (119)$$

or

$$m\langle\vec{v}\rangle(\vec{r},t) = -i\hat{p}\,\varphi(r,t) - e\vec{A}(\vec{r},t) \qquad (120)$$

So, mean velocity $\langle\vec{v}\rangle(\vec{r},t)$ of the probabilistic flow has the scalar potential $\varphi(r,t)$ which is the phase of the wave function $\Psi$. In other words the phase of the wave function sets the velocity potential of the flow of probability.

Interchanging (114), (54) transforms into classical equation of movement for a charged particle in electromagnetic fields $\vec{E}$ and $\vec{B}$. The vortex component of the velocity $\vec{A}$ corresponds with the vector potential of the magnetic field. Equations (56), (57) and conditions (72), (73) coincide with the classical Maxwell equations.

### 7. Conclusion

A mathematically rigorous derivation of the first Vlasov equation as a well-known Schrödinger equation for the probabilistic description of a system and families of the classic diffusion equations and heat conduction for the deterministic description of physical systems was inferred in this paper. A physical meaning of the phase of the wave function which, as it proved, is a scalar potential of the probabilistic flow velocity is demonstrated. In the case when the probabilistic flow velocity has a vortex component, a well-known Pauli equation for one of the spinar components is derived from the Vlasov equation. It is noteworthy that the Schrödinger equation itself was derived according to the phenomenological considerations, and its mathematically rigorous derivation is presented here. A scheme of the construction of the Schrödinger equation solving from the Vlasov equation solving and vice-versa is shown. A process of introduction of the potential to the Schrödinger equation and its interpretation are given. The analysis of the potential properties gives us the Maxwell equation, the equation of the kinematic point movement, and the movement of the medium within electromagnetic fields equation.